\documentclass[runningheads]{llncs}
\usepackage{multirow}
\usepackage{booktabs}
\usepackage{tabularx}
\usepackage{graphicx}
\usepackage{amsfonts}
\usepackage{amsmath}
\usepackage{floatrow}
\usepackage{subfig}

\usepackage[table,xcdraw]{xcolor}

\newcommand{\comment}[1]{}

\begin{document}
\title{Whole Heart Mesh Generation For Image-Based Computational Simulations By Learning Free-From Deformations\thanks{This works was supported by the National Science Foundation, Award No. 1663747.}}
\titlerunning{Whole Heart Mesh Generation For Computational Simulations}
%
\author{Fanwei Kong\inst{1}\orcidID{0000-0003-1190-565X} \and
Shawn C. Shadden\inst{1}\orcidID{0000-0001-7561-1568}}
%
\authorrunning{F.\ Kong \& S.C. Shadden }
%
\institute{Department of Mechanical Engineering, University of California, Berkeley, CA 94720 USA 
\email{\{fanwei\_kong,shadden\}@berkeley.edu}}

\maketitle              
\begin{abstract}
 Image-based computer simulation of cardiac function can be used to probe the mechanisms of (patho)physiology, and guide diagnosis and personalized treatment of cardiac diseases. This paradigm requires constructing simulation-ready meshes of cardiac structures from medical image data--a process that has traditionally required significant time and human effort, limiting large-cohort analyses and potential clinical translations. We propose a novel deep learning approach to reconstruct simulation-ready whole heart meshes from volumetric image data. Our approach learns to deform a  template mesh to the input image data by predicting  displacements  of  multi-resolution control point grids. We discuss the methods of this approach and demonstrate its application to efficiently create simulation-ready whole heart meshes for computational fluid dynamics simulations of the cardiac flow. Our source code is available at \url{https://github.com/fkong7/HeartFFDNet}.
\end{abstract}
\section{Introduction} 
Patient-specific computer models of the heart derived from medical image data have been developed to simulate a variety of aspects of cardiac function, including electrophysiology \cite{Trayanova2011}, hemodynamics \cite{Bavo2016} and tissue mechanics \cite{Marx2020} to explore improvements in cardiovascular diagnoses and treatments \cite{Arevalo2016,ZAHID20161687_AF}, and the biomechanical underpinnings of diseases \cite{Genet_EP_HF,Potse_EP_HF}. However, generating simulation-suitable models of the heart from image data requires significant time and human efforts and is a critical bottleneck limiting clinical applications or large-cohort studies\cite{Mittal,Doost2016}. Deforming-domain computational fluid dynamics (CFD) simulations of the intracardiac hemodynamics, in particular, requires both the geometries and the deformation of the heart from a sequence of image snapshots of the heart throughout the cardiac cycle \cite{Mittal,Vedula2015,KONG2020}.
Challenges of image-based model construction are related to the entwined nature of the heart, difficulty differentiating individual cardiac structures from each other and the surrounding tissue, the large deformations of these structures over the cardiac cycle, as well as complicated steps to label various surfaces or regions for the assignment of boundary conditions or parameters if the model is to be used to support simulations.

Deep learning methods have demonstrated promising performance in automatic whole heart segmentation \cite{ZHUANG2019,Payer2018}. However, prior studies have not generally focused on learning to predict surface meshes directly from medical data for the purpose of computational simulations. Prior efforts on accelerating cardiac model construction for simulations have typically adopted a multistage approach whereby 3D segmentations of cardiac structures are first obtained from image volumes using convolutional neural networks (CNN), meshes of the segmented regions are then generated using marching cube algorithms, and finally manual surface post-processing or editing is performed~\cite{KONG2020,Maher2019}. However, CNN based methods may produce segmentations that achieve high average voxel-wise accuracy but contain extraneous regions that are anatomically inconsistent, unphysical and unintelligible for simulation-based analyses. Correcting such artifacts generally requires a number of carefully designed post-processing steps \cite{KONG2020} that cannot be easily adapted to generating complicated whole heart models. 

Recently, a few studies have sought to directly construct anatomical structures in the form of surface meshes from medical image data \cite{Voxel2Mesh,DeepOrganNet,kong2021deeplearning,Attar2019,Ecabert2008}. Particularly, \cite{kong2021deeplearning} leverages a graph convolutional neural network to predict deformation on mesh vertices from a pre-defined mesh template to fit multiple anatomical structures in a 3D image volume. \cite{Attar2019} uses deep neural networks and patient metadata to directly predict cardiac shape parameters of a statistical shape model of the heart. However, the shape and topology of the predicted meshes are pre-determined by the mesh template and cannot be easily changed to accommodate various mesh requirements for different cardiac simulations. 
In contrast to learning deformation on a template mesh, \cite{DeepOrganNet} proposed to learn the \textit{space} deformation by predicting the displacements of a control point grid to deform template meshes of the lung. However, this approach leveraged memory-intensive, fully-connected neural network layers to predict the displacements with a small number of control points from a 2D X-Ray image and thus cannot be directly applied to model complex whole-heart geometries with large geometric variations from 3D image volumes.

Therefore, we are motivated to automatically and directly generate meshes that are suitable for computational simulations of cardiac function. Our method learns the multi-resolution B-spline free-form deformation of the space to produce detailed whole heart meshes from volumetric CT images by a novel graph convolutional module and feature sampling method. After learning the space deformation, our approach is not limited by a single template and is able to deform mesh templates that have arbitrary resolutions or contains various subsets of cardiac structures. When applied on time-series image data, our approach is able to generate temporally consistent meshes that capture the motion of a beating heart and are suitable for cardiac hemodynamics simulations.

\section{Methods} 
\begin{figure}
\centering
\includegraphics[width=\textwidth]{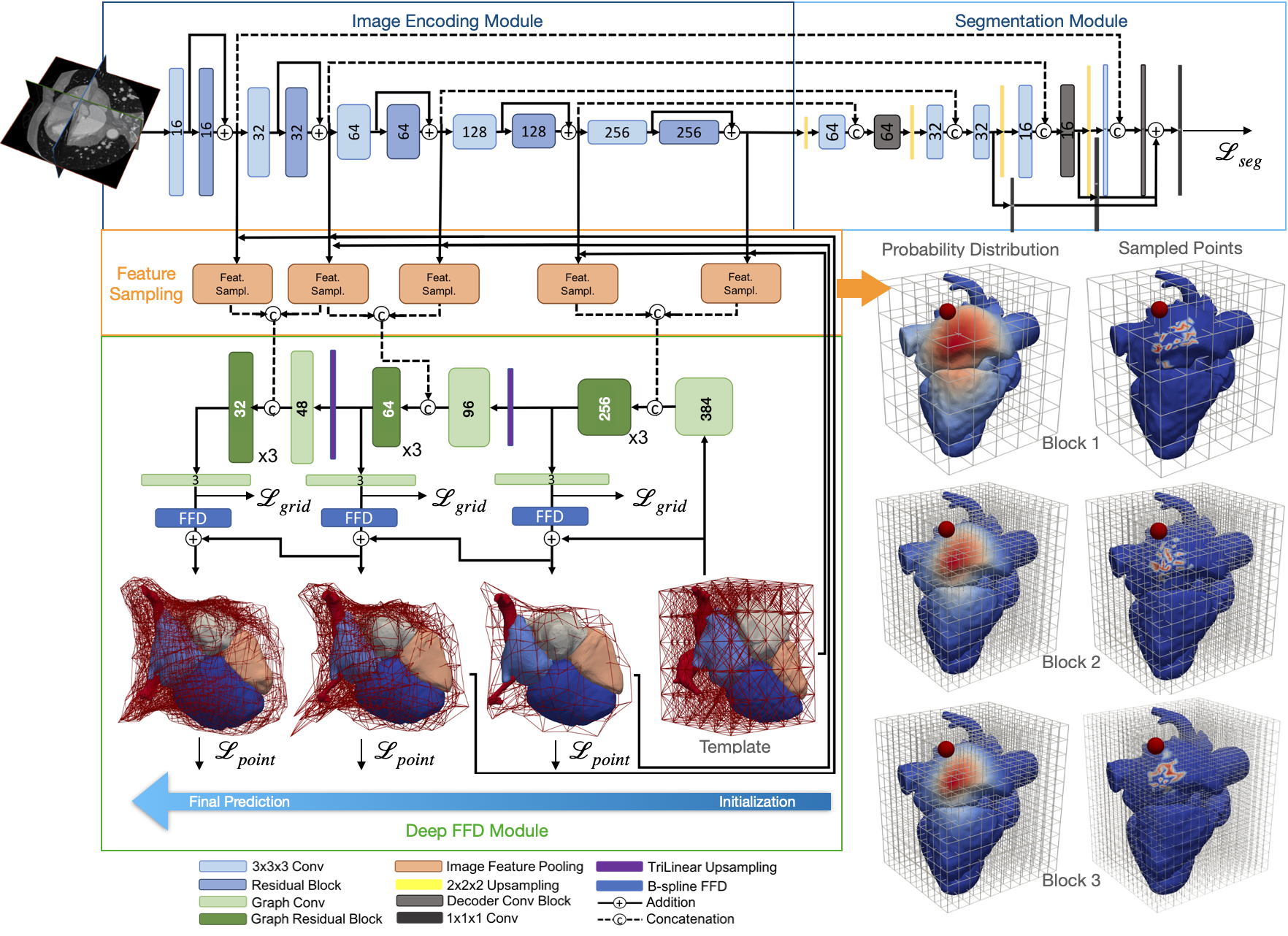}   
\caption{Diagram of the proposed automatic whole heart reconstruction approach.} 
\label{figure:network}
\end{figure}

Fig~\ref{figure:network} shows our proposed whole-heart mesh generation pipeline. Our framework consists of three components to predict the whole-heart meshes from a volumetric input image: (1) an image encoding module, (2) a image feature sampling module, (3) a deep free-form deformation module that predicts control point displacements to deform the template mesh and (4) a segmentation module that predicts a binary segmentation map to allow additional supervision using ground truth annotations.
\subsubsection{B-Spline Based FFD}
We used a 3D tensor product of the 1D cubic B-spline formulation to deform the space. Namely, for a control point gird of $(l+1)\times(m+1)\times(n+1)$, the relation between a deformed mesh vertex $\mathbf{v}$ and the control points $\mathbf{p}$ is described by $ \mathbf{v}(s, t, u) = \sum_{i=0}^l \sum_{j=0}^m \sum_{k=0}^n B_{i,3}(s)B_{j, 3}(t)B_{k,3}(u)\mathbf{p}_{i, j, k}$. $B_{i, 3}$ is the cubic B-spline basis.
Such relation can be expressed in the matrix form $\mathbf{V} = \mathbf{B}\mathbf{P}$, $\mathbf{V}\in \mathbb{R}^{N\times3},  \mathbf{B}\in\mathbb{R}^{N\times\psi},  \mathbf{P}\in\mathbb{R}^{\psi\times3}$, where $N$ and $\psi$ are the number of mesh vertices and control points, respectively. $\mathbf{B}$ is the trivariate B-spline tensor and can be pre-computed from the template mesh. $\mathbf{P}$ is the control point coordinates that the network will learn to predict. Compared with the Bernstein deformation tensor implemented in \cite{DeepOrganNet}, the B-spline-based deformation matrix is sparse since the B-spline basis are defined locally and thus can greatly reduce the computational cost for high-resolution control point grids. 
\subsubsection{Deep FFD Module}
Since the heart involves complicated geometries and significant shape variations during the cardiac cycle and across patients, a dense control point grid is necessary to produce accurate reconstruction of the cardiac structures. Therefore, flattening the image feature vectors of a 3D image and using fully connected layers to predicted control point displacements as proposed in \cite{DeepOrganNet} is no longer computationally feasible. We therefore propose to use a graph convolutional network (GCN) to predict the control point displacements based on sampled image feature vectors. 

\textit{Graph Convolution On Control Grid}:~ 
We represent the control point grid as a graph $\mathcal{M} = (\mathcal{V}, \mathcal{E})$. 
Each control point is connected with all its 26 neighbors (7, 11, and 17 neighbors at the corner point, edge point, and surface point, respectively). 
The graph convolution on a mesh follows \cite{Defferrard2016} and \cite{kong2021deeplearning}. Briefly, we used a first-order Chebyshev polynomial approximation described as $f_{out} = \sigma(\theta_0 f_{in} + \theta_1 f_{in} \Tilde{L} )$, where $\theta_0, \theta_1 \in \mathbb{R}^{d_{out}\times d_{in}}$ are trainable weights, $f_{in} \in \mathcal{R}^{d_{in}\times N}$, $f_{out} \in \mathcal{R}^{d_{out}\times N}$ are input and output feature matrices of a graph convolution layer applied on the control point grid, respectively, and $\Tilde{L} = 2L_{norm}/\lambda_{max} - I$, $\Tilde{L} \in \mathcal{R}^{N\times N}$is the scaled and normalized Laplacian matrix \cite{Defferrard2016}. $N$ is the number of control points. $d_{in}$ and $d_{out}$ are the input and output graph feature dimensions, respectively. The feature lengths of the intermediate layers match with the numbers in Fig~\ref{figure:network}, with 3 for displacements. Compared with conventional convolution with trainable  filters, graph convolution requires far fewer parameters as the connection among vertices is encoded in the graph Laplacian matrix. 

\textit{Deep Multi-Resolution FFD}:~
Our proposed graph decoding module consists of three deformation blocks to progressively deform the template mesh. For the initial mesh deformation blocks, we used lower resolution control point grids conditioned on the more abstracted, high-level image feature maps while using high-resolution control point grids with low-level, high-resolution feature maps for the later mesh deformation blocks. Within each deformation block, we concatenate the sampled image feature vector with the vertex feature vectors on the control points and then use residual graph convolutional blocks to predict the displacements on the control points to deform the template meshes. Between two deformation blocks, we used used trilinear interpolation to upsample the features on lower-resolution control point grid to the same grid resolution as in the next deformation block. The numbers of control points along each dimension were 6, 12, and 16, respectively for the 3 deformation blocks.

\textit{Probability Sampling of Image Features}:~ 
Effective sampling of the image features is  essential for training a dense volumetric feature encoder. 
We randomly sample 16 points on the whole heart per control point based on a normal distribution centered at each control points with the covariance determined by the grid resolution (Fig~\ref{figure:network}). In each FFD block, we update the coordinates of sampled points based on FFD, sample image features at these coordinates and then compute the expectation of image features over the sampled points for each grid point. These image features are then concatenated with the grid features for displacement prediction using GCN.
Control points in the low-resolution grid thus have a larger field of selection than those in the high-resolution grid. Fig~\ref{figure:network} visualizes the probability distribution and sampled points correspond to one control point from control grid at different resolutions. 

\subsubsection{Loss functions}
The training of our networks was supervised by 3D ground truth meshes of the whole heart as well as a binary segmentation indicating occupancy of the heart on the voxel grid that corresponds to the input image volume. 
The total loss is a weighted combination of a point loss, a grid elasticity loss and a segmentation loss over all deformation blocks ${B_i}$. Namely, $\mathcal{L}_{total} = \sum_b^3 \mathcal{L}_{point}(\mathbf{P}^{B_i}, \mathbf{G}^{B_i}) + \alpha_1 \sum_b^3 \mathcal{L}_{grid}(\Delta \mathbf{C}^{B_i}) + \alpha_2 \mathcal{L}_{seg}(I_p, I_g)$. We used $\alpha_1=100$ selected from 10, 100, 1000 based on the validation accuracy. $\alpha_2$ was initially set to 200 (selected from 10, 100 and 200) and decreased by 5\% every 5 epochs during training.
We use the Chamfer loss as the point loss $
    \mathcal{L}_{point} (\mathbf{P}_i, \mathbf{G}_i) =\sum_{\substack{\mathbf{p}\in \mathbf{P}_i}} \min_{\substack{\mathbf{g}\in \mathbf{G}_i}} ||\mathbf{p}-\mathbf{g}||_2^2 + \sum_{\substack{\mathbf{g}\in \mathbf{G}_i}} \min_{\substack{\mathbf{p}\in \mathbf{P}_i}} ||\mathbf{p}-\mathbf{g}||_2^2 
$, where $\mathbf{p}$ and $\mathbf{g}$ are, respectively, points from vertex sets of the predicted mesh $\mathbf{P}_i$ and the ground truth mesh $\mathbf{G}_i$ of cardiac structure $i$.
Since excessive deformation of the control points especially during early phase of training may introduce undesirable mesh artifacts, we use the grid elasticity loss to regularize the network to predict small displacements of the control points, $\mathcal{L}_{grid}(\Delta \mathbf{C}) = \sum_{\substack{\mathbf{c}\in \mathbf{C}}} ||\Delta\mathbf{c} - \frac{1}{N}\sum_{\substack{\mathbf{c}\in \mathbf{C}}}\Delta\mathbf{c}||_2^2 $. 
We used a hybrid loss function $\mathcal{L}_{seg}(I_p, I_g)$ that sums the cross-entropy and the dice losses between the predicted occupancy probability map $I_p$ and the ground truth binary segmentation of the whole heart $I_g$. The validation loss converged in 36 hrs on a GTX1080Ti GPU.

\subsubsection{Cardiac flow simulation}
We applied the Arbitrary Lagrangian-Eulerian (ALE) formulation of the incompressible Navier-Stokes equations to simulate the intraventricular flow and account for deforming volumetric mesh using the finite element method. The volumetric mesh was created automatically from our FFD predicted surface mesh using TetGen \cite{Si2015}. 
Blood was assumed to have a viscosity $\mu$ of $4.0 \times 10^{-3} Pa \cdot s$ and a density $\rho$ of $1.06 g/cm^3$.
The equations were solved with the open-source svFSI solver from the SimVascular project \cite{Updegrove2016}.

\section{Experiments and Results} 
\subsubsection{Dataset and Preprocessing}
We applied our method to public datasets  of contrast-enhanced CT images from both normal and abnormal hearts and mostly cover the whole hearts, MMWHS~\cite{ZHUANG2019}, orCalScore~\cite{orCaScore} and SLAWT~\cite{SLAWT}. Intensity normalization and resizing as well as data augmentation techniques, random scaling, rotation, shearing and elastic deformation were applied following the procedures in \cite{kong2021deeplearning}. The training and validation datasets contained 87 and 15 CT images, respectively. The 40 CT images from MMWHS test dataset and 10 sets of times-series CT data \cite{kong2021deeplearning} were left out for evaluations. The ground truth labels include the 4 heart chambers, aorta, pulmonary artery, parts of the pulmonary veins and venae cavae for the training and validation data. 

\begin{figure}[h]
\centering
\includegraphics[width=0.95\textwidth]{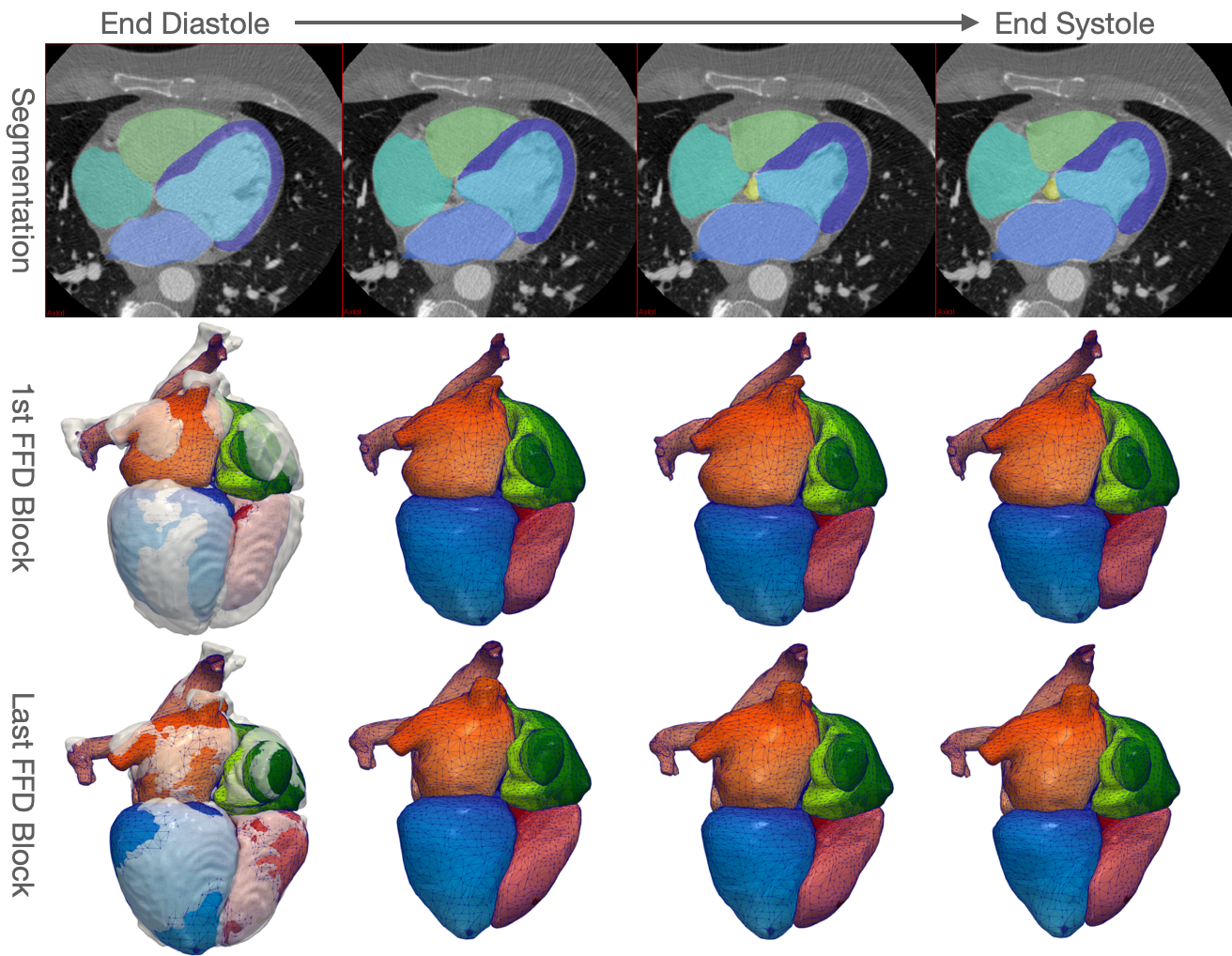} 
\caption{Whole-heart reconstruction results for time-series CT data. The first row shows predicted segmentation overlaid with CT images. The second and third rows shows mesh predictions from the first and the last deep FFD blocks. The predictions at the first time frame are overlaid with ground truths. Color maps denotes the mesh vertex IDs.} 
\label{figure:corr}
\end{figure}

\subsubsection{Generation of 4D Meshes for CFD Simulations}
We applied our method on time-series CT image data that consisted of images from 10 time frames over the cardiac cycle for each patient. Fig~\ref{figure:corr} compares the predictions from the first and last FFD blocks. A low-resolution control point grid can capture the general shape and location of the heart in the image stack whereas the high-resolution control point grid can capture further detailed features. From the segmentation results in Fig~\ref{figure:corr}, our method is able to capture the minor changes between time frames. Furthermore, as denoted by the color maps of vertex IDs, our method consistently deforms the template meshes such that predictions across different time frames have feature correspondence.

\begin{figure}[h]
\centering
\includegraphics[width=0.95\textwidth]{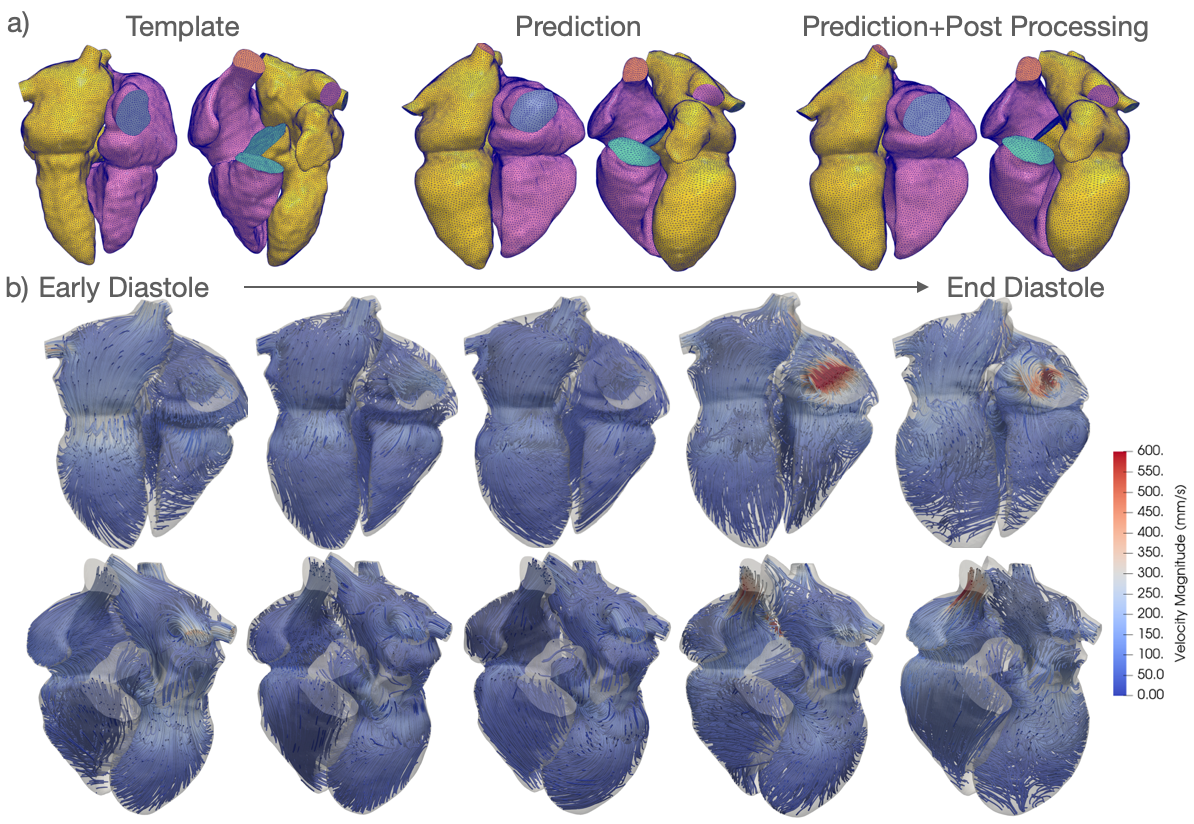}   
\caption{a) Simulation-ready templates and example predictions. b) CFD simulation results using the predicted 4D meshes.} 
\label{figure:simulation}
\end{figure}

We also evaluated the potential of our method to generate simulation-ready meshes for deforming-domain CFD simulations of cardiac flow. We used a simulation ready template with trimmed inlet/outlet geometries and tagged face IDs for prescribing boundary conditions. Fig~\ref{figure:simulation}a shows the template mesh and our prediction for a representative patient. As our method does not constrain the faces of inlets or outlets to be co-planer, we post-processed our prediction using automatic scripts to project the points on the tagged inlet and outlet faces to fitted planes. We used the resulting meshes to simulate the filling phase of heart after interpolating the 4D meshes to increase the temporal resolution to 0.001s. For the fluid domain, Dirichlet (displacements) boundary conditions were applied on the chamber walls as well as on aorta and pulmonary outlets, while Neumann (pressure) boundaries conditions were applied on pulmonary vein and vena cava inlets. Fig~\ref{figure:simulation}b displays the simulation results of the velocity streamlines at multiple time steps during diastole. Videos of the predicted meshes and simulation results of more cases are in our supplementary materials.  

\subsubsection{Comparison of different methods}
We compared the whole heart reconstruction performance of different FFD strategies, namely, 1) using the same resolution of control point grid for all deformation blocks as in \cite{DeepOrganNet} and uniformly sample the image feature space based the the coordinates of the control points (Single-Res + US), 2) using multi-resolution control point grids and uniformly sample the image feature space based on the coordinates of the high resolution grid (Multi-Res + US) and 3) our final model that uses multi-resolution control point grids and probability sampling of the image feature space based on the coordinates of the whole heart template (Multi-Res + WHS). Our supplementary materials additionally include an ablation study of individual loss components and the use of GCN decoder. Furthermore, we compared these FFD-based methods with prior whole-heart reconstruction or segmentation methods, Kong et al \cite{kong2021deeplearning}, 2DUNet \cite{Ronneberger2015,KONG2020}, residual 3D UNet \cite{Isensee2019AnAA} and Voxel2Mesh\cite{Voxel2Mesh}. We followed procedures described in \cite{kong2021deeplearning} to implement those methods. 

\begin{table}
\centering
\caption{A comparison of prediction accuracy on MMWHS CT test datasets from different deep FFD methods.}\label{table:mmwhs}
\resizebox{\textwidth}{!}{%
\begin{tabular}{cccccccccc}
\toprule
       &               &             Epi &              LA &              LV &              RA &              RV &               Ao &               PA &               WH \\
\midrule
 \multirow{3}{*}{\parbox{1.2cm}{Dice}} & Singe-Res+US &   0.72$\pm$0.09 &   0.82$\pm$0.08 &   0.81$\pm$0.07 &   0.79$\pm$0.07 &   0.81$\pm$0.06 &    0.78$\pm$0.09 &    0.69$\pm$0.14 &    0.79$\pm$0.05 \\
            & Multi-Res+US &   0.81$\pm$0.06 &   0.89$\pm$0.05 &   0.88$\pm$0.07 &   0.84$\pm$0.07 &   0.86$\pm$0.05 &    0.87$\pm$0.07 &    0.76$\pm$0.13 &    0.86$\pm$0.04 \\
            & Multi-Res+WHS &   \textbf{0.84$\pm$0.05} &   \textbf{0.91$\pm$0.04} &   \textbf{0.89$\pm$0.07} &   \textbf{0.86$\pm$0.06} &   \textbf{0.88$\pm$0.04} &    \textbf{0.91$\pm$0.04} &      \textbf{0.8$\pm$0.1} &    \textbf{0.88$\pm$0.03} \\
\cline{1-10}
    \multirow{3}{*}{\parbox{1.2cm}{ASSD (mm)}} & Singe-Res+US &   2.28$\pm$0.76 &   2.61$\pm$0.97 &   2.72$\pm$1.02 &   2.87$\pm$0.93 &   2.42$\pm$0.56 &    2.41$\pm$0.96 &    3.55$\pm$1.62 &    2.69$\pm$0.69 \\
            & Multi-Res+US &   1.61$\pm$0.43 &    1.5$\pm$0.56 &   1.62$\pm$0.66 &   2.09$\pm$0.92 &    1.59$\pm$0.4 &    1.29$\pm$0.52 &     2.5$\pm$1.38 &    1.75$\pm$0.41 \\
            & Multi-Res+WHS &   \textbf{1.41$\pm$0.38} &    \textbf{1.4$\pm$0.47} &   \textbf{1.46$\pm$0.68} &   \textbf{1.87$\pm$0.84} &   \textbf{1.49$\pm$0.42} &    \textbf{1.08$\pm$0.35} &    \textbf{2.19$\pm$1.17} &    \textbf{1.54$\pm$0.34} \\
\cline{1-10}
    \multirow{3}{*}{\parbox{1.2cm}{HD (mm)}} & Singe-Res+US &  15.44$\pm$2.42 &  11.54$\pm$3.47 &  10.45$\pm$3.13 &  15.22$\pm$5.35 &   12.02$\pm$2.5 &    14.9$\pm$7.32 &  26.41$\pm$11.49 &  27.87$\pm$10.59 \\
            & Multi-Res+US &  14.45$\pm$2.49 &   9.25$\pm$2.92 &   8.04$\pm$2.25 &  13.55$\pm$5.84 &  10.86$\pm$2.59 &   12.54$\pm$5.53 &  \textbf{23.97$\pm$12.62} &  \textbf{25.67$\pm$11.51} \\
            & Multi-Res+WHS &  \textbf{13.51$\pm$2.59} &   \textbf{8.58$\pm$2.87} &   \textbf{7.66$\pm$2.61} &  \textbf{12.75$\pm$5.46} &  \textbf{10.09$\pm$2.48} &   \textbf{12.24$\pm$6.86} &  24.79$\pm$12.52 &  26.76$\pm$11.17 \\
\bottomrule
\end{tabular}
}
\end{table}

\begin{figure}
\centering
\includegraphics[width=0.95\textwidth]{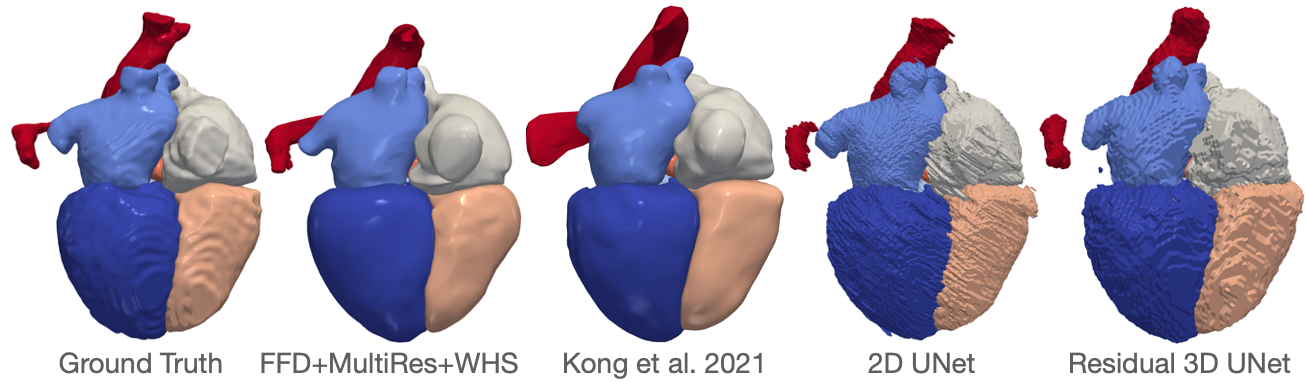} 
\caption{Qualitative comparisons among different methods} 
\label{figure:compare}
\end{figure}

Table \ref{table:mmwhs} presents the accuracy scores of different FFD-based methods evaluated on the MMWHS test dataset. Since the ground truths of MMWHS test dataset do not contain the pulmonary veins or venae cavae, we used a template without those structures to generate our predictions. For FFD-based methods, Multi-Res consistently produced more accurate geometries for all cardiac structures (Table \ref{table:mmwhs}). Multi-Res + WHS produced more accurate geometries with less surface artifacts than Multi-Res + US, indicating the contribution of our proposed sampling method. Fig~\ref{figure:compare} displays qualitative whole-heart reconstruction results from different methods. Mesh-deformation based methods, ours and \cite{kong2021deeplearning}, are able to generate smoother and more anatomically consistent geometries compared with segmentation-based approaches, which produced surfaces with staircase artifacts and disconnected regions (Fig~\ref{figure:compare}). However, Kong, et al., 2021 produced overly smoothed pulmonary veins and vena cava geometries, likely because these are elongated structures and that method deforms spheres rather than a more fitting template of each structure as used in our method here. Quantitatively, our method is generally able to produce similar geometric accuracy compared with prior state-of-the-art methods, while having the additional advantage of directly support various cardiac simulations. Nonetheless, we did observe slightly reduced level of accuracy (see supplementary materials), likely because it is challenging to use a single whole-heart template to fully capture the geometric variations across patients. In ongoing work we plan to add a template retrieval module to automatically select a template that best suits the input.

\section{Conclusion}
We proposed a novel deep-learning approach to directly construct whole heart meshes from image data. We learn to deform a template mesh to match the input data by predicting displacements of multi-resolution control point grids. To our knowledge, this is the first approach that is able to directly generate whole heart meshes for computational simulations and allows switching template meshes to accommodate different modeling requirements. We demonstrated application of our method on constructing a dynamic whole heart mesh from time-series CT image data to simulate the cardiac flow driven by the cardiac motion. Our method was able to construct such meshes within a minute on a standard desktop computer (3 GHz Intel Core i5 CPU) whereas prior methods can take hours of time and human efforts to generate simulation-ready 4D meshes.

%
%
\bibliographystyle{splncs04}
\bibliography{ref}

\title{— Supplementary material — \\ Whole Heart Mesh Generation For Image-Based Computational Simulations By Learning Free-From Deformations\thanks{This works was supported by the National Science Foundation, Award No. 1663747.}}
\titlerunning{Whole Heart Mesh Generation For Computational Simulations}
%
\author{Fanwei Kong\inst{1}\orcidID{0000-0003-1190-565X} \and
Shawn C. Shadden\inst{1}\orcidID{0000-0001-7561-1568}}
%
\authorrunning{F.\ Kong \& S.C. Shadden }
%
\institute{Department of Mechanical Engineering, University of California, Berkeley, CA 94720 USA 
\email{\{fanwei\_kong,shadden\}@berkeley.edu}}
%
\maketitle      
\begin{table}
\centering
\caption{Ablation study: a comparison of prediction accuracy on MMWHS CT test dataset after removing grid loss $L_{grid}$, removing segmentation loss $L_{seg}$, and using conventional convolution with $3\times3\times$ filters rather than graph convolution in the decoder. Removing $L_{grid}$ or $L_{seg}$ as well as using a classical $3\times3\times$ convolution resulted in drop of geometric accuracy of the whole heart reconstruction.}\label{table:ablation}
\resizebox{\textwidth}{!}{%
\begin{tabular}{cccccccccc}
\toprule
       &               &             Epi &              LA &              LV &              RA &              RV &               Ao &               PA &               WH \\
\midrule
\multirow{4}{*}{Dice} & Final FFDNet &   0.839 &   0.905 &   0.894 &   0.863 &   0.877 &   0.906 &   0.802 &   0.878 \\
            & CNN Decoder &   0.806 &   0.872 &   0.874 &   0.826 &   0.846 &   0.843 &   0.764 &   0.845 \\
            & No Grid Loss &   0.772 &   0.889 &   0.855 &   0.833 &   0.850 &   0.879 &   0.772 &   0.843 \\
            & No Segmentation Loss &   0.781 &   0.884 &   0.859 &   0.837 &   0.851 &   0.874 &   0.780 &   0.845 \\
\cline{1-10}
    \multirow{4}{*}{Jaccard} & Final FFDNet &   0.725 &   0.829 &   0.814 &   0.764 &   0.783 &   0.831 &   0.680 &   0.784 \\
            & CNN Decoder &   0.679 &   0.777 &   0.781 &   0.710 &   0.737 &   0.736 &   0.628 &   0.734 \\
            & No Grid Loss &   0.636 &   0.804 &   0.755 &   0.720 &   0.743 &   0.790 &   0.642 &   0.731 \\
            & No Segmentation Loss &   0.647 &   0.796 &   0.761 &   0.726 &   0.745 &   0.783 &   0.651 &   0.734 \\
\cline{1-10}
    \multirow{4}{*}{ASSD (mm)} & Final FFDNet &   1.406 &   1.404 &   1.455 &   1.870 &   1.488 &   1.080 &   2.193 &   1.544 \\
            & CNN Decoder &   1.709 &   1.782 &   1.668 &   2.281 &   1.846 &   1.650 &   2.470 &   1.898 \\
            & No Grid Loss &   1.899 &   1.574 &   2.009 &   2.165 &   1.770 &   1.257 &   2.599 &   1.904 \\
            & No Segmentation Loss &   1.816 &   1.638 &   1.865 &   2.084 &   1.711 &   1.250 &   2.473 &   1.832 \\
\cline{1-10}
    \multirow{4}{*}{HD (mm)} & Final FFDNet &  13.508 &   8.584 &   7.665 &  12.753 &  10.089 &  12.243 &  24.794 &  26.765 \\
            & CNN Decoder &  14.076 &   9.713 &   7.694 &  14.037 &  10.905 &  12.886 &  24.533 &  26.413 \\
            & No Grid Loss &  14.330 &   9.521 &  10.086 &  14.163 &  12.013 &  12.607 &  25.361 &  27.688 \\
            & No Segmentation Loss &  14.268 &   9.122 &   9.154 &  13.567 &  10.792 &  13.408 &  23.862 &  26.036 \\
\bottomrule
\end{tabular}
}
\end{table}

\begin{table}
\centering
\caption{Robustness validation of our proposed FFDNet under different initialization and data splits. The table displays the mean geometric accuracy measures and their standard deviations on the MMWHS CT test dataset for our proposed FFDNet trained using 5 different random initialization and training/validation data splits.}
\resizebox{\textwidth}{!}{%
\begin{tabular}{ccccccccc}
\toprule
&               Epi &                LA &               LV &                RA &                RV &                Ao &                PA &                WH \\
\midrule
Dice &    0.835$\pm$0.004 &   0.905$\pm$0.003 &  0.895$\pm$0.001 &    0.86$\pm$0.004 &   0.878$\pm$0.005 &   0.904$\pm$0.004 &   0.813$\pm$0.007 &   0.878$\pm$0.002 \\
   Jaccard &     0.72$\pm$0.005 &   0.829$\pm$0.004 &  0.815$\pm$0.002 &    0.76$\pm$0.005 &   0.786$\pm$0.007 &   0.827$\pm$0.008 &    0.697$\pm$0.01 &   0.784$\pm$0.004 \\
   ASSD (mm) &    1.459$\pm$0.036 &   1.404$\pm$0.035 &  1.439$\pm$0.015 &   1.932$\pm$0.043 &   1.468$\pm$0.043 &   1.005$\pm$0.054 &   2.077$\pm$0.102 &   1.536$\pm$0.031 \\
   HD (mm) &   13.954$\pm$0.264 &    8.61$\pm$0.132 &  7.647$\pm$0.014 &  12.971$\pm$0.226 &  10.447$\pm$0.228 &  11.753$\pm$0.758 &  23.439$\pm$0.815 &  25.592$\pm$0.722 \\
\bottomrule
\end{tabular}
}
\end{table}

\begin{table}
\centering
\caption{Quantitative comparison of the geometric accuracy of the reconstruction results between the proposed method and prior methods, \cite{kong2021deeplearning}, \cite{Ronneberger2015}, \cite{Isensee2019AnAA} and \cite{Voxel2Mesh} on MMWHS CT test dataset. Detailed implementation of the prior methods has been described in \cite{kong2021deeplearning} and we used the same image pre-processing and training data for all methods.  Our method is able to provide similar or slightly reduced level of accuracy compared with prior methods while having the additional advantage of directly support various cardiac simulations.}\label{table:ablation}
\resizebox{\textwidth}{!}{%
\begin{tabular}{cccccccccc}
\toprule
           &            &     Epi &      LA &      LV &      RA &      RV &      Ao &      PA &      WH \\
\midrule
\multirow{5}{*}{Dice} & FFDNet &   0.839 &   0.905 &   0.894 &   0.863 &   0.877 &   0.906 &   0.802 &   0.878 \\
            & Kong \textit{et al.} \cite{kong2021deeplearning}&   0.899 &   0.932 &   0.940 &   0.892 &   0.910 &   0.950 &   0.852 &   0.918 \\
            & 2DUNet \cite{Ronneberger2015} &   0.899 &   0.931 &   0.931 &   0.877 &   0.905 &   0.934 &   0.832 &   0.911 \\
            & 3DUNet \cite{Isensee2019AnAA} &   0.863 &   0.902 &   0.923 &   0.868 &   0.876 &   0.923 &   0.813 &   0.888 \\
            & Voxel2Mesh \cite{Voxel2Mesh} &   0.775 &   0.888 &   0.910 &   0.857 &   0.885 &   0.874 &   0.758 &   0.865 \\
\cline{1-10}
    \multirow{5}{*}{Jaccard} & FFDNet &   0.725 &   0.829 &   0.814 &   0.764 &   0.783 &   0.831 &   0.680 &   0.784 \\
            & Kong \textit{et al.} \cite{kong2021deeplearning}&   0.819 &   0.875 &   0.888 &   0.809 &   0.837 &   0.905 &   0.755 &   0.849 \\
            & 2DUNet \cite{Ronneberger2015}&   0.817 &   0.872 &   0.873 &   0.787 &   0.828 &   0.879 &   0.726 &   0.837 \\
            & 3DUNet \cite{Isensee2019AnAA}&   0.762 &   0.825 &   0.861 &   0.769 &   0.783 &   0.860 &   0.695 &   0.799 \\
            & Voxel2Mesh \cite{Voxel2Mesh}&   0.638 &   0.801 &   0.839 &   0.754 &   0.795 &   0.778 &   0.619 &   0.763 \\
\cline{1-10}
    \multirow{5}{*}{ASSD (mm)} & FFDNet &   1.406 &   1.404 &   1.455 &   1.870 &   1.488 &   1.080 &   2.193 &   1.544 \\
            & Kong \textit{et al.} \cite{kong2021deeplearning}&   1.335 &   1.042 &   0.842 &   1.583 &   1.176 &   0.531 &   1.904 &   1.213 \\
            & 2DUNet \cite{Ronneberger2015}&   0.808 &   1.049 &   0.905 &   1.719 &   1.064 &   0.645 &   1.551 &   1.088 \\
            & 3DUNet \cite{Isensee2019AnAA}&   1.443 &   1.528 &   1.024 &   1.943 &   1.663 &   0.814 &   2.194 &   1.552 \\
            & Voxel2Mesh \cite{Voxel2Mesh}&   1.714 &   1.696 &   1.266 &   2.020 &   1.492 &   1.341 &   3.398 &   1.848 \\
\cline{1-10}
    \multirow{5}{*}{HD (mm)} & FFDNet &  13.508 &   8.584 &   7.665 &  12.753 &  10.089 &  12.243 &  24.794 &  26.765 \\
            & Kong \textit{et al.} \cite{kong2021deeplearning}&  14.393 &  10.407 &  10.325 &  13.639 &  13.360 &   9.407 &  26.616 &  28.035 \\
            & 2DUNet \cite{Ronneberger2015}&   9.980 &   8.773 &   6.098 &  13.624 &  10.016 &  10.013 &  27.834 &  28.727 \\
            & 3DUNet \cite{Isensee2019AnAA}&  13.635 &  10.814 &   9.580 &  16.031 &  15.635 &  13.326 &  26.941 &  31.088 \\
            & Voxel2Mesh \cite{Voxel2Mesh}&  13.564 &   8.743 &   6.248 &  12.116 &   9.601 &  12.080 &  26.252 &  27.459 \\
\bottomrule
\end{tabular}
}
\end{table}
\end{document}